\documentclass{eptcs}
\providecommand{\event}{NCL 2022} 
\usepackage{breakurl}             
\usepackage{underscore}           

\usepackage{xcolor}

\usepackage[utf8]{inputenc}
\usepackage{graphicx}
\usepackage{xcolor}
\usepackage{appendix}
\usepackage{hyperref}

\usepackage{proof}
\usepackage{stmaryrd}
\usepackage[all]{xy}
\usepackage{amssymb}
\usepackage{amsmath}
\usepackage{amsthm}
\usepackage[inline]{enumitem}
\usepackage[normalem]{ulem}
\usepackage{wasysym}
\usepackage{multicol}

\theoremstyle{definition}

\newtheorem*{example}{Example}

\newcommand{\doi}[1]{\href{https://doi.org/#1}{doi: #1}}

\newcommand{\eqdf}{\mathbin{=_{\mathrm{df}}}}

\newcommand{\fresh}{\mathsf{fresh}}
\newcommand{\Set}{\mathsf{Set}}

\newcommand{\I}{\mathsf{I}}
\newcommand{\ot}{\otimes}
\newcommand{\ol}{\mathbin{\diagup}}
\newcommand{\lo}{\mathbin{\diagdown}}

\newcommand{\loe}{{\mathsf{E}}_{\mathbin{{\multimap}}}}
\newcommand{\loi}{{\mathsf{I}}_{\mathbin{{\multimap}}}}

\newcommand{\rmst}{\mathsf{rmst}}
\newcommand{\lmst}{\mathsf{lmst}}

\newcommand{\rcstol}{\mathsf{rcst}_{{\ol}}}
\newcommand{\rcstlo}{\mathsf{rcst}_{{\lo}}}
\newcommand{\lcstol}{\mathsf{lcst}_{{\ol}}}
\newcommand{\lcstlo}{\mathsf{lcst}_{{\lo}}}
\newcommand{\nattrans}{\stackrel{\cdot}{\to}}

\newcommand{\otE}{E_{{\ot}}}
\newcommand{\otET}{E_{{\ot}}^T}
\newcommand{\otI}{I_{{\ot}}}
\newcommand{\IE}{E_{{\I}}}
\newcommand{\IET}{E_{{\I}}^T}
\newcommand{\II}{I_{{\I}}}
\newcommand{\loE}{E_{{\lo}}}
\newcommand{\loI}{I_{{\lo}}}
\newcommand{\olE}{E_{{\ol}}}
\newcommand{\olI}{I_{{\ol}}}
\newcommand{\ax}{\mathsf{ax}}
\newcommand{\sw}{\mathsf{sw}_{{\dn}}}

\newcommand{\bE}{E_{{!}}}
\newcommand{\bET}{E_{{!}}^T}
\newcommand{\bI}{I_{{!}}}

\newcommand{\Ip}{\widehat{\I}}
\newcommand{\otp}{\mathbin{\widehat{\ot}}}
\newcommand{\olp}{\mathbin{\widehat{\ol}}}
\newcommand{\lop}{\mathbin{\widehat{\lo}}}
\newcommand{\bp}{\mathbin{\widehat{!}}}

\newcommand{\fmasem}[1]{\llbracket #1 \rrbracket}
\newcommand{\cxtsem}[1]{\llbracket #1 \rrbracket}
\newcommand{\tmsem}[1]{\llbracket #1 \rrbracket}
\newcommand{\tsem}{\tmsem}
\newcommand{\envsem}[1]{\llbracket #1 \rrbracket}
\newcommand{\reify}{\downarrow}
\newcommand{\reflect}{\uparrow}

\newcommand{\dn}{\Downarrow}
\newcommand{\up}{\Uparrow}

\newcommand{\Cxt}{\mathsf{Cxt}}
\newcommand{\LCxt}{\mathsf{LCxt}}
\newcommand{\ICxt}{\mathsf{ICxt}}

\newcommand{\At}{\mathsf{At}}

\newcommand{\Tm}{\mathsf{Fma}}
\newcommand{\Fma}\Tm
\newcommand{\id}{\mathsf{id}}

\newcommand{\sub}{\mathsf{cut}}
\newcommand{\cut}{\sub}
\newcommand{\subst}{\mathsf{sub}}

\newcommand{\refl}{\mathsf{refl}}

\newcommand{\run}{\mathsf{run}}
\newcommand{\runup}{\mathsf{run}^{{\up}}}

\newcommand{\emb}{\mathsf{emb}}

\newcommand{\nbe}{\mathsf{nbe}}

\newcommand{\proofbox}[1]{\begin{tabular}{c} #1 \end{tabular}}

\newcommand{\ids}{\mathsf{ids}}

\renewcommand{\L}{\textbf{L}}
\newcommand{\Lup}{\ensuremath{\L_{\beta\eta}}}

\newcommand{\tr}{\triangleright}
\newcommand{\nil}{\langle\rangle}

\newcommand{\deq}{\sim}
\newcommand{\ceq}{\approx}

\newcommand{\embup}{\emb_{\up}}
\newcommand{\embdn}{\emb_{\dn}}

\title{Normalization by Evaluation for the Lambek Calculus}
\author{
  Niccol{\`o} Veltri
\institute{Tallinn University of Technology, Tallinn, Estonia}
\email{niccolo@cs.ioc.ee}}
\def\titlerunning{Normalization by Evaluation for the Lambek Calculus}
\def\authorrunning{N. Veltri}

\begin{document}
\maketitle              
\begin{abstract}
  The syntactic calculus of Lambek is a deductive system for the
  multiplicative fragment of intuitionistic non-commutative linear
  logic. As a fine-grained calculus of resources, it has many
  applications, mostly in formal computational investigations of natural languages.
  
  This paper introduces a calculus of $\beta\eta$-long normal forms for
  derivations in the Lambek calculus with multiplicative unit and conjunction among its logical connectives. Reduction to normal form follows the
  normalization by evaluation (NbE) strategy: (i) evaluate a
  derivation in a Kripke model of Lambek calculus;
  (ii) extract normal forms from the obtained semantic values. The
  implementation of the NbE algorithm requires the presence of a
  strong monad in the Kripke interpretation of positive
  formulae, in analogy with the extension of intuitionistic propositional logic with falsity and disjunction. The NbE algorithm can also be instantiated with minor variations to calculi for related substructural logics, such as multiplicative and dual intuitionistic linear logic (MILL and DILL).


\end{abstract}

\section{Introduction}
\label{sec:intro}

The syntactic calculus \L\ of Lambek~\cite{Lam:matss} is a deductive
system which is primarily employed in mathematical studies of sentence
structure in natural language. From a logical perspective, it provides
a proof system for the multiplicative fragment of intuitionistic
non-commutative linear logic~\cite{Abr:noncil,PP:natdin}, comprising
only of two ordered linear implications (or residuals) $\ol$ and $\lo$,
tensor product $\ot$ and (often but not always) a unit $\I$ as logical
connectives.

The metatheory of the Lambek calculus has been thoroughly developed in
the past decades, in particular its categorical semantics by Lambek
himself~\cite{Lam:ded1,Lam:ded2}. The Lambek calculus enjoys cut
elimination~\cite{Lam:matss} and various normalization procedures,
e.g. by Hepple for the implicational fragment~\cite{Hep:norftp} or more
recently by Amblard and Retor{\'e}~\cite{AR:norspl}, aimed at the
reduction of the proof search space and consequently the number of
possible derivation of a given sequent. Various diagrammatic calculi
and proof nets for the Lambek calculus have also been
proposed~\cite{Roo:pronlc,LR:pronlc}.

In this work, we study the natural deduction presentation of the
Lambek calculus \L, together with a calculus \Lup\ consisting of
$\beta\eta$-long normal forms, i.e. derivations that do not contain any
redexes, and no further $\eta$-expansion is applicable. Sequents in
\Lup\ have two shapes: $\Gamma \up A$, consisting of derivations in
$\beta\eta$-long normal form, and $\Gamma \dn A$, consisting of
neutral derivations, i.e. (under Curry-Howard correspondence with a
non-commutative linear variant of typed $\lambda$-calculus) a variable
applied to other normal forms. The design of \Lup\ is inspired by the
intercalation calculus for (non-commutative) linear
logic~\cite{BNS:focnd}. It appears in particular as a fragment of the
calculus of normal forms for ordered linear
logic introduced by Polakov and Pfenning~\cite{PP:natdin}.

The normalization algorithm, sending each derivation in \L\ to its
$\beta\eta$-normal form in \Lup, is an instance of \emph{normalization
  by evaluation} (NbE)~\cite{BS:inveft,AHS:catrrf}. NbE is a
well-established normalization procedure for many variants of (typed
and untyped) $\lambda$-calculus. The main idea behind NbE is that the
effective normalization procedure factors through a categorical model
of the syntactic calculus, usually a Kripke/presheaf model. Each
formula $A$ can be interpreted as an object $\fmasem{A}$ of the model,
and this interpretation can be extended to contexts $\cxtsem{\Gamma}$
as well. Moreover, each derivation $t$ of a sequent $\Gamma \vdash A$
in \L, for which we employ the syntax $t : \Gamma \vdash A$, is
\emph{evaluated} to a map $\tmsem{t} : \cxtsem{\Gamma}
\nattrans \fmasem{A}$. The Kripke model is precisely chosen so that
the semantic value $\tmsem{t}$ contains enough information to allow
the extraction of a normal form in \Lup. This extraction process,
sending semantic values to normal forms, is typically called
\emph{reification}. The presence of mixed-variance connectives, such
as the ordered implications $\lo$ and $\ol$, requires the construction
of the reification function to be defined simultaneously with a
\emph{reflection} function, embedding neutrals in the Kripke
model. The NbE procedure consists in the implementation of a function
$\nbe : \Gamma \vdash A \to \Gamma \up A$ whose construction can be
schematized as follows:
\[
\arraycolsep=1pt
\begin{array}{ccccc}
  (\text{syntactic derivation}) & & (\text{Kripke semantic value}) &
  & (\beta\eta\text{-long normal form}) \\
  t : \Gamma \vdash A & \longmapsto & \tmsem{t} : \cxtsem{\Gamma} \nattrans \fmasem{A} &
  \longmapsto & \nbe\;t : \Gamma \up A \\
  &  \mathbf{evaluation} & & \mathbf{reification} & 
\end{array}
\]

Normalization by evaluation for the implicational fragment of \L, not
including unit and tensor, can be extrapolated from the normalization
algorithm for the implicational fragment of ordered linear logic
investigated by Polakov in his PhD thesis~\cite{PP:natdin,Pol:ordlla}. Similarly,
this could also be reconstructed from the recently developed NbE
procedure for the natural deduction calculus of skew prounital closed
categories~\cite{UVZ:dedscs}.

Nevertheless, the presence of positive logical connectives $\I$ and
$\ot$ in combination with the negative implications $\lo$ and $\ol$
does not make the NbE procedure for the implicational fragment of
\L\ directly extensible to the full calculus including also rules for
$\I$ and $\ot$. This situation sheds many similarities with the case
of intuitionistic propositional logic with falsity and disjunction
(or, equivalently, typed $\lambda$-calculus with empty type and sum
types), where the standard Kripke semantics proves itself to be too
weak for the extraction of normal forms. In the literature, the
situation has been fixed in two different ways: employing a
\emph{strong monad} in the presheaf model for the interpretation of
sums \cite{AU:nore,AS:norecp,VR:expebc} or switching to more
convoluted sheaf-theoretic models \cite{ADHS:noretl,BCF:extntd}.  In
this paper we follow the first solution and similarly extend our
presheaf model with a strong monad for recovering the reification
procedure. This solution only helps with the implementation of NbE
wrt. to a $\beta\eta$-conversion that does not include all possible
permutative conversions, e.g. in the calculus \Lup\ the
$\ot$-introduction and $\ot$-elimination rules do not commute.
Defining NbE for a stronger conversion, removing all possible
nondeterministic choices during proof search, is left to future work
(more discussion on this in the conclusive section).

The main difference between NbE for the Lambek calculus \L\ and intuitionistic propositional logic (or typed $\lambda$-calculus) is the extra bureaucracy that the substructural system \L, in which all the structural rules of weakening, contraction and exchange are absent, requires wrt. hypotheses/resources in context. Nevertheless, the presence or absence of the exchange rule does not seem to play a fundamental role. In fact, in the final part of the paper, we briefly discuss how to adapt NbE to the case of multiplicative intuitionistic linear logic (MILL)~\cite{BBdePH:tercil}. We also see a further extension to dual intuitionistic linear logic (DILL), which is a particular presentation of MILL extended with a linear exponential modality !~\cite{BP:duaill}.

The material in the paper is organized as follows.
Section~\ref{sec:natded} introduces the Lambek calculus \L\ in natural
deduction. Section~\ref{sec:weak} presents the calculus \Lup\ of
$\beta\eta$-long normal forms. Section~\ref{sec:nbe} discusses the NbE
procedure: the construction of the Kripke model and the strong monad
on it (Section~\ref{sec:kripke}), the interpretation of syntactic
constructs (Section~\ref{sec:inter}) and the reification/reflection
algorithms for computing normal forms (Section~\ref{sec:nbefun}).
Section~\ref{sec:corr} proves the correctness of NbE using logical
relations. Section~\ref{sec:milldill} discusses extensions of NbE to MILL and DILL.

\section{The Lambek Calculus in Natural Deduction}
\label{sec:natded}

The Lambek calculus \L\ has formulae generated by the grammar $A,B ::= p
\ | \ \I \ | \ A \ot B \ | \ B \ol A \ | \ A \lo B$, where $p$ comes from a
given set $\At$ of atomic formulae, $\I$ is the multiplicative unit, $\ot$ is a multiplicative
conjunction (a.k.a. tensor product), while $\ol$ and $\lo$ are left and
right implications (a.k.a. left and right residuals). Sequents in \L\ are
pairs $\Gamma \vdash A$, where $\Gamma$ is an ordered (possibly empty)
list of formulae, that we call context, and $A$ is a single
formula. The sets of formulae and contexts are called $\Fma$ and $\Cxt$,
respectively. Derivations in \L\ are generated by the inference
rules in Figure~\ref{eq:natded}.
We write $t : \Gamma \vdash A$ to
indicate that $t$ is a particular derivation of $\Gamma \vdash A$. 
\begin{figure}[t!]
\begin{equation*}
  \def\arraystretch{1.4}
  \begin{array}{c}
  \infer[\olI]{\Gamma \vdash B \ol A}{
    \Gamma, A \vdash B
  }  
  \qquad
  \infer[\loI]{\Gamma \vdash A \lo B}{
    A, \Gamma \vdash B
  }
  \qquad
  \infer[\olE]{\Gamma,\Delta \vdash B}{
    \Gamma \vdash B \ol A & \Delta \vdash A
  }
  \qquad
  \infer[\loE]{\Gamma,\Delta \vdash B}{
    \Gamma \vdash A & \Delta \vdash A \lo B
  }
  \\
    \infer[\ax]{A \vdash A}{}
    \qquad
  \infer[\II]{~ \vdash \I}{}
  \qquad
  \infer[\IE]{\Delta_0 , \Gamma , \Delta_1 \vdash C}{
    \Gamma \vdash \I & \Delta_0 , \Delta_1 \vdash C
  }
  \\
  \infer[\otI]{\Gamma,\Delta \vdash A \ot B}{
    \Gamma \vdash A & \Delta \vdash B
  } 
  \qquad
  \infer[\otE]{\Delta_0 , \Gamma , \Delta_1 \vdash C}{
    \Gamma \vdash A \ot B & \Delta_0 , A,B,\Delta_1 \vdash C
  }
\end{array}
\end{equation*}
\caption{Inference rules of the Lambek calculus \L\ in natural deduction}
\label{eq:natded}
\end{figure}

Substitution, i.e. the \emph{cut rule}, is admissible in
\L~\cite{Lam:matss}. As is common in the NbE literature, our cut rule allows the simultaneous substitution of multiple variables in context:  
\begin{equation}\label{eq:cut}
  \infer[\sub]{\Gamma \vdash A}{
    \Gamma \tr \Delta & \Delta \vdash A
  }
\end{equation}
where $\Gamma \tr \Delta$ denotes a set containing lists of derivations, called
\emph{environments}: given $\Gamma = \Gamma_1,\dots,\Gamma_n$ and
$\Delta = \langle A_1,\dots,A_n \rangle$, an element of $\Gamma \tr
\Delta$ consists of derivations of sequents $\Gamma_i \vdash A_i$, for
all $i = 1,\dots,n$. The relation $\tr$ can be defined inductively by
the rules:
\begin{equation*}
\infer[\nil]{\nil \tr \nil}{
}
\qquad\qquad
\infer[\mathsf{cons}]{\Gamma,\Gamma' \tr A,\Delta}{
\Gamma \vdash A & \Gamma' \tr \Delta}
\end{equation*}
In other words, the only environment in $\nil \tr \nil$ is the empty
list of terms, denoted also by $\nil$. An environment in
$\Gamma,\Gamma' \tr A,\Delta$ is a pair of a term $t : \Gamma \vdash
A$ and another environment $\sigma : \Gamma' \tr \Delta$. We always
write the pair as $(t,\sigma)$ instead of the more verbose
$\mathsf{cons}\;t\,\sigma$.  More generally, given two environments
$\sigma_1 : \Gamma_1 \tr \Delta_1$ and $\sigma_2 : \Gamma_2 \tr
\Delta_2$, we write $(\sigma_1,\sigma_2) : \Gamma_1, \Gamma_2 \tr
\Delta_1, \Delta_2$ for their concatenation. Given a context $\Gamma$,
recursively define the lists of variables $\ids_\Gamma : \Gamma \tr
\Gamma$ by induction on $\Gamma$:
  $\ids_{\nil} \eqdf \nil$ and 
  $\ids_{A,\Gamma} \eqdf (\ax,\ids_\Gamma)$.
A more familiar-looking cut rule is derivable from (\ref{eq:cut}) as follows:
\[
\proofbox{
  \infer[\subst]{\Delta_0,\Gamma,\Delta_1 \vdash A}{
    t : \Gamma \vdash C &
    u : \Delta_0,C,\Delta_1 \vdash A
  }
}
\eqdf
\proofbox{
  \infer[\cut]{\Delta_0,\Gamma,\Delta_1 \vdash A}{
    (\ids_{\Delta_0},t,\ids_{\Delta_1}) : \Delta_0,\Gamma,\Delta_1 \tr \Delta_0,C,\Delta_1 &
    u : \Delta_0,C,\Delta_1 \vdash A
  }
}
\]
The employment of $\cut$ in place of $\subst$ in NbE facilitates the
statement and proof of correctness discussed in
Section~\ref{sec:corr}.

\begin{figure}[t!]
\begin{equation*}\arraycolsep=1.4pt
  \begin{array}{rcl@{\hspace{14pt}}r}
    \multicolumn{4}{l}{\textsc{($\beta$-conversions)}}\\
    \olE\;(\olI\;t)\;u & \deq & \subst\;u\;t
    & ( t : \Gamma,A \vdash B,\ u : \Delta \vdash A) \\
    \loE\;u\;(\loI\;t) & \deq & \subst\;u\;t
    & ( u : \Gamma \vdash A,\ t : A,\Delta \vdash B) \\
        \IE\;\II\;t & \deq & t
    & ( t : \Delta_0,\Delta_1 \vdash C)\\
        \otE\;(\otI\;s_1\;s_2)\;t & \deq & \subst\;s_1\;(\subst\;s_2\;t)
        &  ( s_i : \Gamma_i \vdash A_i,\ t : \Delta_0,A_1,A_2,\Delta_1 \vdash C)
        \\[.6em]
    \multicolumn{4}{l}{\textsc{($\eta$-conversions)}}\\
    t &\deq& \olI\;(\olE\;t\;\ax)
    & ( t : \Gamma \vdash B \ol A) \\
    t &\deq& \loI\;(\loE\;\ax\;t)
    & ( t : \Gamma \vdash A \lo B)\\
    s &\deq& \IE\;s\;\II
    & ( s : \Gamma \vdash \I) \\
    s &\deq& \otE\;s\;(\otI\;\ax\;\ax)
    & ( s : \Gamma \vdash A \ot B) \\[.6em]
    \multicolumn{4}{l}{\textsc{(permutative conversions)}} \\
    \IE\;s\;(\olI\;t) &\deq& \olI\;(\IE\;s\;t) &
    (s : \Gamma \vdash \I,\ t : \Delta_0,\Delta_1,A\vdash B)
    \\
    \IE\;s\;(\loI\;t) &\deq& \loI\;(\IE\;s\;t) &
      (s : \Gamma \vdash \I,\ t : A,\Delta_0,\Delta_1\vdash B)
    \\
    \olE\;(\IE\;s\;t)\;u &\deq& \IE\;s\;(\olE\;t\;u) &
    ( s : \Gamma \vdash \I,\ t : \Delta_0,\Delta_1 \vdash B \ol A,\
    u : \Omega \vdash A)
    \\
    \olE\;t\;(\IE\;s\;u) &\deq& \IE\;s\;(\olE\;t\;u) &
    ( s : \Gamma \vdash \I,\ t : \Omega \vdash B \ol A,\
    u : \Delta_0,\Delta_1 \vdash A)
    \\
    \loE\;(\IE\;s\;t)\;u &\deq& \IE\;s\;(\loE\;t\;u) &
    ( s : \Gamma \vdash \I,\ t : \Delta_0,\Delta_1 \vdash A ,\
    u : \Omega\vdash A \lo B)
    \\
    \loE\;t\;(\IE\;s\;u) &\deq& \IE\;s\;(\loE\;t\;u) &
    (s : \Gamma \vdash \I,\ t : \Omega \vdash A,\
    u : \Delta_0,\Delta_1 \vdash A\lo B)
    \\
    \IE\;(\IE\;s_1\;s_2)\;t &\deq& \IE\;s_1\;(\IE\;s_2\;t) &
    (s_1 : \Gamma \vdash \I,\ s_2 : \Delta_0,\Delta_1\vdash \I,\
    t : \Omega_0,\Omega_1 \vdash C) \\
    \otE\;s\;(\olI\;t) &\deq& \olI\;(\otE\;s\;t) &
    (s : \Gamma \vdash A'\ot B',\ t : \Delta_0,A',B',\Delta_1,A\vdash B)
    \\
    \otE\;s\;(\loI\;t) &\deq& \loI\;(\otE\;s\;t) &
      (s : \Gamma \vdash A'\ot B',\ t : A,\Delta_0,A',B',\Delta_1\vdash B)
    \\
    \olE\;(\otE\;s\;t)\;u &\deq& \otE\;s\;(\olE\;t\;u) &
    ( s : \Gamma \vdash A'\ot B',\ t : \Delta_0,A',B',\Delta_1 \vdash B \ol A,\
    u : \Omega \vdash A)
    \\
    \olE\;t\;(\otE\;s\;u) &\deq& \otE\;s\;(\olE\;t\;u) &
    ( s : \Gamma \vdash A'\ot B',\ t : \Omega \vdash B \ol A,\
    u : \Delta_0,A',B',\Delta_1 \vdash A)
    \\
    \loE\;(\otE\;s\;t)\;u &\deq& \otE\;s\;(\loE\;t\;u) &
    ( s : \Gamma \vdash A'\ot B',\ t : \Delta_0,A',B',\Delta_1 \vdash A ,\
    u : \Omega\vdash A \lo B)
    \\
    \loE\;t\;(\otE\;s\;u) &\deq& \otE\;s\;(\loE\;t\;u) &
    (s : \Gamma \vdash A'\ot B',\ t : \Omega \vdash A,\
    u : \Delta_0,A',B',\Delta_1 \vdash A\lo B)
    \\
    \otE\;(\otE\;s_1\;s_2)\;t &\deq& \otE\;s_1\;(\otE\;s_2\;t) &
    (s_1 : \Gamma \vdash A\ot B,\ s_2 : \Delta_0,A,B,\Delta_1\vdash A' \ot B',\
    t : \Omega_0,A',B',\Omega_1 \vdash C)
  \end{array}
\end{equation*}
\caption{The $\beta\eta$-equivalence of derivations in \L}
\label{eq:eqs1}
\end{figure}

Derivations in \L\ can be identified modulo a certain $\beta\eta$-equivalence relation $\deq$, which is
the least congruence generated by the pairs of derivations in
Figure~\ref{eq:eqs1}. Equations include $\beta$- and
$\eta$-conversions for all the logical connectives. Due to space
limitations, derivations in these equations are displayed using an inline,
term-like notation, but hopefully the reader will not struggle
too hard in reconstructing the associated proof trees. E.g. the reconstructed $\beta$-rules for right implication $\ol$ and multiplicative conjunction $\ot$ look are:
\[
\proofbox{
  \infer[\olE]{\Gamma,\Delta \vdash B}{
    \infer[\olI]{\Gamma \vdash B \ol A}{
      t : \Gamma,A \vdash B
    }
    &
    u : \Delta \vdash A
  }
}
\deq
\proofbox{
  \infer[\subst]{\Gamma,\Delta \vdash B}{
    u : \Delta \vdash A
    &
    t : \Gamma ,A \vdash B
  }
}
\]
\[\small
\proofbox{
  \infer[\otE]{\Delta_0,\Gamma_1,\Gamma_2,\Delta_1 \vdash C}{
    \infer[\otI]{\Gamma_1,\Gamma_2 \vdash A_1 \ot A_2}{
      s_1 : \Gamma_1 \vdash A_1
      &
      s_2 : \Gamma_2 \vdash A_2
    }
    & \hspace{-2mm}
    t : \Delta_0,A_1,A_2,\Delta_1 \vdash C
  }
}\hspace{-3mm}
\deq
\proofbox{
  \infer[\subst]{\Delta_0,\Gamma_1,\Gamma_2,\Delta_1 \vdash C}{
    s_1 : \Gamma_1 \vdash A_1
    & \hspace{-3mm}
    \infer[\subst]{\Delta_0,A_1,\Gamma_2,\Delta_1 \vdash C}{
      s_2 : \Gamma_2 \vdash A_2
      & \hspace{-1mm}
      t : \Delta_0,A_1,A_2,\Delta_1 \vdash C
    }
  }
}
\]
Equations in Figure~\ref{eq:eqs1} include also many permutative conversions, but not all of them. For example, the following equations are missing, stating that $\ot$-introduction commutes with $\ot$-elimination, and that two $\ot$-eliminations can be swapped if they are independent, i.e. they act on disjoint collections of formulae in the context:
\begin{equation}\label{eq:missing}
\arraycolsep=1.4pt
  \begin{array}{rcl@{\qquad}r}
    \otI\;(\otE\;s\;t)\;u &\deq& \otE\;s\;(\otI\;t\;u) &
    (s : \Gamma \vdash A' \ot B',\ t : \Delta_0,A',B',\Delta_1 \vdash A,\
    u : \Omega \vdash B) \\
    \otI\;t\; (\otE\;s\;u) &\deq& \otE\;s\;(\otI\;t\;u) &
    (s : \Gamma \vdash A' \ot B',\ t : \Omega \vdash A,\
    u : \Delta_0,A',B',\Delta_1 \vdash B) \\
    \otE\;s\;(\otE\;s'\;t) & \deq& \otE\;s'\;(\otE\;s\;t) &
    (s : \Gamma \vdash A \ot B,\ s' : \Gamma' \vdash A' \ot B',\
    t : \Delta_0,A,B,\Delta_1,A',B',\Delta_2 \vdash C)
  \end{array}
\end{equation}
E.g. the proof trees in the first equation of (\ref{eq:missing}) are:
\[\small
\proofbox{
  \infer[\otI]{\Delta_0,\Gamma,\Delta_1,\Omega \vdash A \ot B}{
    \infer[\otE]{\Delta_0,\Gamma,\Delta_1 \vdash A}{
      s: \Gamma \vdash A' \ot B'
      &
      t : \Delta_0,A',B',\Delta_1 \vdash A
    }
    &
    u : \Omega \vdash B
    }
}\hspace{-3mm}
\deq
\proofbox{
  \infer[\otE]{\Delta_0,\Gamma,\Delta_1,\Omega \vdash A \ot B}{
    s: \Gamma \vdash A' \ot B'
    & \hspace{-1mm}
    \infer[\otI]{\Delta_0,A',B',\Delta_1,\Omega \vdash A \ot B}{
      t : \Delta_0,A',B',\Delta_1 \vdash A
      &
      u : \Gamma' \vdash B
    }
  } 
}
\]
Analogous permutative conversions involving $\otI$ and $\IE$, two independent applications of $\IE$, and an application of $\IE$ independent from an application of $\otE$, are missing as well.
The study of the Lambek calculus with derivations identified by a richer system of equations, including the missing axioms discussed above, such as the ones in (\ref{eq:missing}),  is left for future work. Categorically speaking, the equational theory resulting from adding all the missing axioms corresponds to the one of monoidal biclosed categories. The Lambek calculus \L, with derivations quotiented by the resulting richer congruence relation, would therefore be a presentation of the free monoidal biclosed category on the set $\At$.



\section{A Calculus of Normal Forms}
\label{sec:weak}

Canonical representatives of equivalence classes for the congruence
$\deq$, i.e. $\beta\eta$-long normal forms, can be organized in a
calculus called \Lup.
Sequents in \Lup\ have two shapes, $\Gamma \up A$ and $\Gamma \dn
A$. We employ the notation using ${\up}$ and ${\dn}$ from
\cite{PP:natdin}, which is more generally used in focused sequent
calculi for linear logic~\cite{BNS:focnd}. In the literature on intuitionistic
propositional logic (i.e. typed $\lambda$-calculus, via
Curry-Howard correspondence), derivations of $\Gamma \up A$ are called
\emph{normal forms}, while derivations of $\Gamma \dn A$ are called
\emph{neutrals}. A calculus with similar backward- and
forward-chaining phases is also typically called an
\emph{intercalation calculus}.

Derivations in \Lup\ are generated by the 
inference rules in Figure~\ref{eq:weak}.
The metavariable $\gamma^+$ indicates a \emph{non-negative} formula, i.e. a formula which is not of the form $A \lo B$ nor $B \ol A$. Elimination rules $\IE$ and $\otE$ are only applicable when the goal formula $\gamma^+$ in the conclusion is non-negative.

\begin{figure}[t!]
\begin{equation*}
  \def\arraystretch{1.4}
  \begin{array}{c}
  \infer[\olI]{\Gamma \up B \ol A}{
    \Gamma, A \up B
  }  
  \qquad
  \infer[\loI]{\Gamma \up A \lo B}{
    A, \Gamma \up B
  }
  \qquad
  \infer[\olE]{\Gamma,\Delta \dn B}{
    \Gamma \dn B \ol A & \Delta \up A
  }
  \qquad
  \infer[\loE]{\Gamma,\Delta \dn B}{
    \Gamma \up A & \Delta \dn A \lo B
  }
    \\
    \infer[\ax]{A \dn A}{}
    \qquad
      \infer[\II]{~ \up \I}{
  }
  \qquad
  \infer[\IE]{\Delta_0 , \Gamma , \Delta_1 \up \gamma^+}{
    \Gamma \dn \I & \Delta_0 , \Delta_1 \up \gamma^+
  }
  \\
    \infer[\otI]{\Gamma,\Delta \up A \ot B}{
    \Gamma \up A & \Delta \up B
  }
  \qquad
  \infer[\otE]{\Delta_0 , \Gamma , \Delta_1 \up \gamma^+}{
    \Gamma \dn A \ot B & \Delta_0 , A,B,\Delta_1 \up \gamma^+
  }
  \qquad
    \infer[\sw]{\Gamma \up p}{
    \Gamma \dn p}
\end{array}
\end{equation*}
\caption{Inference rules of the calculus \Lup\ of $\beta\eta$-long normal forms}
\label{eq:weak}
\end{figure}

When restricting the attention on the implicational fragment of \Lup, with only one implication $\lo$, it is possible to recognize in the rules of Figure~\ref{eq:weak} a goal-directed proof search strategy attempting to build a derivation in \L. The construction of a valid derivation would proceed as follows: first eagerly apply the $\lo$-introduction rule, until the succedent becomes atomic; then switch to the neutral phase $\dn$ and apply a sequence of $\lo$-eliminations rules to arguments in normal form, ultimately closing the derivation using the $\ax$ rule. Under the Curry-Howard correspondence between \L\ and a non-commutative linear (some people might say ``planar''~\cite{UVZ:dedscs}) variant of $\lambda$-calculus, the derivations in this implicational fragment of \Lup\ correspond to weak head normal forms. The rules for the other implication $\ol$ have analogous roles. 

The introduction and elimination rules of $\ot$ and $\I$ both produce normal forms, but the first premises of $\otE$ and $\IE$ are neutrals. The fact that all the rules involving the positive connectives $\ot$ and $\I$ are in the same phase $\up$ shows that permutative conversions such as those in (\ref{eq:missing}) do not hold (as syntactic equalities) in \Lup.
The formula $\gamma^+$ appearing in the rules $\otE$ and $\IE$ is required to be different from an implication, which fixes the relative position of $\lo$- and $\ol$-introduction wrt. the elimination rules for the positive connectives $\ot$ and $\I$.

Soundness of \Lup\ wrt. \L\ is evident: each \Lup-derivation can be embedded
into \L\ via functions $\embup : \Gamma \up A \to \Gamma \vdash A$ and
$\embdn : \Gamma \dn A \to \Gamma \vdash A$, which simply change $\up$
and $\dn$ to $\vdash$ and erase all uses of the rule $\sw$.

\section{Normalization by Evaluation}
\label{sec:nbe}

\Lup\ is also complete wrt. \L. The proof of completeness corresponds
to the construction of a normalization algorithm $\nbe$, taking a
derivation of $\Gamma \vdash A$ and returning a derivation of $\Gamma
\up A$, satisfying the three following properties, for all derivations
$t,u : \Gamma \vdash A$ and $n : \Gamma \up A$:
\begin{enumerate}
  \item $t \deq u \to \nbe \;t = \nbe \;u$, which means that the
    normalization algorithm sends $\deq$-related derivations in \L\ to
    syntactically equal derivations in \Lup;
  \item $t \deq \embup \;(\nbe\;t)$, which means that each derivation
    in \L\ is $\deq$-related to (the embedding of) its normal form;
  \item $\nbe\;(\embup\;n) = n$, which implies that each derivation in
    \Lup\ corresponds uniquely to an equivalence class of the relation $\deq$ in \L.
\end{enumerate}
The procedure $\nbe$ is defined following the \emph{normalization by
evaluation (NbE)} methodology~\cite{BS:inveft,AHS:catrrf}, consisting in the following steps: $(i)$ Find a model of \L\ and its
equational theory $\deq$, in our case (and in the majority of
applications of NbE) a Kripke/presheaf model. This provides the existence of an
element $\tmsem{t}$ in the model, for each derivation $t$ in \L, such
that $\tmsem{t} = \tmsem{u}$ whenever $t \deq u$; $(ii)$ Define a
reification function, sending a semantic value in the Kripke
model to a normal form in \Lup, so that $\nbe\;t$ is defined as the
reification of $\tmsem{t}$.

\subsection{The Kripke Model}\label{sec:kripke}
The model is defined as the presheaf category $\Set^\Cxt$. Explicitly, an object
$P$ of the category $\Set^\Cxt$ is a $\Cxt$-indexed family of sets,
i.e. a presheaf: for any context $\Gamma$, $P\;\Gamma$ is a
set\footnote{We think of $\Cxt$ as a discrete category, which
  is why there is no mention of $P$'s action of maps.}. A
map $f$ between $P$ and $Q$ in $\Set^\Cxt$ is a $\Cxt$-indexed
family of functions, i.e. a natural transformation: for any context
$\Gamma$, $f\;\Gamma$ is a function between $P\;\Gamma$ and $Q\;\Gamma$. We
typically omit the index $\Gamma$, and simply write $f : P\;\Gamma \to
Q\;\Gamma$. The set of maps between $P$ and $Q$ is denoted $P
\nattrans Q$.

The category $\Set^\Cxt$ is \emph{monoidal biclosed} (or, using Lambek terminology, \emph{residual})~\cite{Lam:ded1}, with
unit, tensor and internal homs given by
\begin{equation}\label{eq:monbi}
  \begin{array}{rcl}
  \Ip \;\Gamma & \eqdf & (\Gamma = \nil) \\
(P \otp Q) \; \Gamma & \eqdf & \{\Gamma_0,\Gamma_1 : \Cxt\} \times \{ \Gamma = \Gamma_0 ,
  \Gamma_1 \} \times P\;\Gamma_0 \times Q\;\Gamma_1 \\
  (P \olp Q) \; \Gamma & \eqdf & \{\Delta :\Cxt\} \to Q\;\Delta \to P\;(\Gamma,\Delta) \\
    (Q \lop P) \; \Gamma & \eqdf & \{\Delta :\Cxt\} \to Q\;\Delta \to P\;(\Delta,\Gamma)  
  \end{array}
\end{equation}
The unit $\Ip\;\Gamma$ is the singleton set which contains an element
if and only if $\Gamma$ is empty. The tensor $(P \otp Q) \; \Gamma$
consists of pairs of an element of $P \; \Gamma_0$ and an element of
$Q\;\Gamma_1$, for some contexts $\Gamma_0$ and $\Gamma_1$ such that
$\Gamma = \Gamma_0,\Gamma_1$. The tensor product $\otp$ is often
called \emph{Day convolution}. The left (resp. right) internal hom $(Q
\lop P) \; \Gamma$ (resp. $(P \olp Q) \; \Gamma$) consists of
functions from $Q\;\Delta$, for a given context $\Delta$, to
$P\;(\Delta,\Gamma)$ (resp. $P\;(\Gamma,\Delta)$).  The tensor and
internal homs in $\Set^\Cxt$ form two adjunctions: there are natural bijective correspondences between the set of maps $P \otp Q \nattrans R$ and the sets of maps $P \nattrans R \olp Q$ and $Q \nattrans P \lop R$.

In (\ref{eq:monbi}) we have employed notation from Agda/Martin L\"of
type theory for the dependent sum and dependent product operations:
$(x : A) \times B\;x$ and $(x : A) \to B\;x$ stand for $\sum_{x :
  A}B\;x$ and $\prod_{x : A}B\;x$ respectively, where $A$ is a set and
$B$ is a family of sets indexed by $A$. Curly brackets indicate
implicit arguments, e.g. when giving an element of $\{x : A\} \times
B\;x$ it is sufficient to give an element $y : B\;x$ for some implicit
$x : A$. For the readers more familiar with set-theoretic notation,
the definitions in (\ref{eq:monbi}) can be rephrased as follows, where for clarity all the appearing arguments have been made explicit:
\begin{equation*}
  \begin{array}{rcl}  
    \Ip\;\Gamma &\eqdf & \{ * \mid \Gamma \text{ is empty} \}
    \\
    (P \otp Q)\;\Gamma &\eqdf & \{(\Gamma_0,\Gamma_1,x,y) \mid \Gamma_0,\Gamma_1 : \Cxt \text{ such that } \Gamma = \Gamma_0,\Gamma_1, \text{ and } x : P\;{\Gamma_0} \text{ and } y : Q\;{\Gamma_1} \} \\
    (P \olp Q)\;\Gamma &\eqdf & \prod_{\Delta : \Cxt} \;Q\;\Delta \to P\;(\Gamma,\Delta) \\
    (Q \lop P)\;\Gamma &\eqdf & \prod_{\Delta : \Cxt} \;Q\;\Delta \to P\;(\Delta,\Gamma)
  \end{array}  
\end{equation*}

The tensor ${\otp}$ is associative and unital wrt. $\Ip$ only up to
natural isomorphism, i.e. $(P \otp Q) \otp S \cong P \otp (Q \otp S)$
and $\Ip \otp P \cong P \cong P \otp \Ip$, and the same is not true if
we replace $\cong$ with $=$. To ease the readability of the
forthcoming constructions, we leave implicit all applications of the natural
isomorphisms of associativity and unitality in the paper.


The monoidal biclosed category $\Set^\Cxt$ is not completely suitable
for the construction of an algorithm satisfying the NbE specification
(more on this in Section~\ref{sec:nbefun}). In analogy with the case of
intuitionistic propositional logic with falsity and
disjunction~\cite{AS:norecp}, we introduce a monad $T$ on
$\Set^\Cxt$. For each presheaf $P$ and context $\Gamma$, the elements
of the set $T\;P\;\Gamma$ are inductively generated by the following
constructors:
\begin{equation}\label{eq:monad}
  \infer[\eta]{T\;P\;\Gamma}{
    P\; \Gamma
  }
  \qquad\qquad
  \infer[\IET]{T\;P\;(\Delta_0,\Gamma,\Delta_1)}{
    \Gamma \dn \I &
    T\;P\;(\Delta_0 , \Delta_1)
  }
  \qquad\qquad
  \infer[\otET]{T\;P\;(\Delta_0,\Gamma,\Delta_1)}{
    \Gamma \dn A \ot B &
    T\;P\;(\Delta_0 , A , B , \Delta_1)
  }
\end{equation}
Notice the similarity of the $\IET$ and $\otET$ constructors with the elimination rules $\IE$ and $\otE$ of \Lup. In other words, elements of $T\;P\;\Gamma$ are lists of neutral terms with a positive formula (i.e. unit or tensor) in the succedent, ending with an element of $P\;\Gamma$.
As it is usual in category theory, we also write $T : P \nattrans Q
\ \to \ T\;P \nattrans T\;Q$ for the action on maps of $T$, and
$\mu : T\;(T\;P) \nattrans T\;P$ for the monad multiplication.
The monad $T$ is left- and right-strong wrt. the monoidal
structure $(\Ip,\otp)$. Left strength is defined by recursion:
\begin{equation}\label{eq:lmst}
  \arraycolsep=1.2pt
  \begin{array}{lcl}
    \multicolumn{3}{l}{\lmst : P \otp T\;Q \ \nattrans \ T \;(P \otp  Q)}\\
    \lmst\;(x,\eta\;y) & \eqdf & \eta\;(x,y) \\
    \lmst\;(x,\IET\;t\;y) & \eqdf & \IET\;t\;(\lmst\;(x,y)) \\
    \lmst\;(x,\otET\;t\;y) & \eqdf & \otET\;t\;(\lmst\;(x,y))
  \end{array}
\end{equation}
Right-monoidal strength $\rmst : T\;P \otp Q \ \nattrans \ T \;(P \otp
Q)$ is defined analogously. The monoidal strengths are interdefinable
with closed strengths $\lcstol : P \olp Q \ \nattrans \ T \;P \olp
T\;Q$ and $\rcstol : T \;(P \olp Q) \ \nattrans \ T\;P \olp Q$, and
also $\lcstlo$ and $\rcstlo$ obtained by turning $\olp$ into $\lop$.
\begin{equation*}
  \arraycolsep=1.2pt
  \begin{array}{lcl}
    \multicolumn{3}{l}{\lcstol : P \olp Q \ \nattrans \ T \;P \olp T\;Q} \\
    \lcstol\;f\;(\eta\;x) &\eqdf& \eta\;(f\;x) \\
    \lcstol\;f\;(\IET\;t\;x) &\eqdf& \IET\;t\;(\lcstol\;f\;x) \\
    \lcstol\;f\;(\otET\;t\;x) &\eqdf& \otET\;t\;(\lcstol\;f\;x)
  \end{array}
  \qquad\qquad
  \begin{array}{lcl}
    \multicolumn{3}{l}{\rcstol : T \;(P \olp Q) \ \nattrans \ T\;P \olp Q} \\
    \rcstol\;(\eta\;f)\;x &\eqdf& \eta\;(f\;x) \\
    \rcstol\;(\IET\;t\;f)\;x &\eqdf& \IET\;t\;(\rcstol\;f\;x)\\
    \rcstol\;(\otET\;t\;f)\;x &\eqdf& \otET\;t\;(\rcstol\;f\;x)
  \end{array}
\end{equation*}
The monad $T$ is reminiscent of a proof-relevant variant of the
closure operator employed in phase semantics of non-commutative
intuitionistic linear logic, appearing in the definition of
non-commutative intuitionistic phase space~\cite{Abr:noncil}.

\subsection{Interpretation of Syntax}\label{sec:inter}
Each formula $A$ is interpreted as a presheaf $\fmasem{A}$:
\begin{equation}\label{eq:fma}
  \begin{array}{c} 
    \fmasem{p}  \eqdf  - \up p \qquad
    \fmasem{\I} \eqdf T\;\Ip \qquad
  \fmasem{A \ot B}   \eqdf   T\;(\fmasem{A} \otp \fmasem{B}) \\
  \fmasem{B \ol A}   \eqdf   \fmasem B \olp \fmasem A \qquad
  \fmasem{A \lo B}   \eqdf    \fmasem A \lop \fmasem B
  \end{array}
\end{equation}
Notice the presence of the monad $T$ in the interpretation of the positive formulae
$\I$ and $\ot$. The interpretation of formulae extends to contexts via the monoidal structure of $\Set^\Cxt$:
  $\cxtsem{\nil} \eqdf \Ip$ and
  $\cxtsem{A, \Gamma} \eqdf \fmasem A \otp \cxtsem{\Gamma}$.
We use the same notation $\fmasem{-}$ for the interpretation of
formulae and contexts (and later for the interpretation of derivations
and environments), but it should always be clear which interpretation function is employed in each situation.


The interpretation of derivations of \L\ in the Kripke model requires
the monad $T$ to be \emph{runnable} on each interpreted formula
$\fmasem{A}$, i.e. we want a natural transformation $\run_A :
T\;\fmasem{A} \nattrans \fmasem{A}$. In turn, this requires
the monad $T$ to be runnable on presheaves of the form $- \up
A$. The function $\run_A$ is defined by induction on $A$, the
interesting cases are $\ol$ and $\lo$, which figure an application of
right closed strengths $\rcstol$ and $\rcstlo$.
\begin{equation}\label{eq:run}
  \begin{array}{lcl}
    \multicolumn{3}{l}{\runup : T\;(- \up A) \nattrans - \up A} \\
    \runup \;(\eta\;t) & \eqdf & t \\
    \runup \; (\IET\;t \;u) & \eqdf & \IE\;t\;(\runup\;u) \\
    \runup \; (\otET\;t \;u) & \eqdf & \otE\;t\;(\runup\;u)
  \end{array}
  \qquad \qquad
  \begin{array}{lcl}
    \multicolumn{3}{l}{\run_A : T\;\fmasem{A} \nattrans \fmasem{A}} \\
    \run_p \;t & \eqdf & \runup \;t \\
    \run_{\I} \;t & \eqdf & \mu\;t \\
    \run_{A \ot B} \; t & \eqdf & \mu\; t \\
    \run_{B \ol A} \; t & \eqdf & \lambda x.\, \run_B \; (\rcstol \;t\;x)\\
    \run_{A \lo B} \; t & \eqdf & \lambda x.\, \run_B \; (\rcstlo \;t\;x)
  \end{array}
\end{equation}

Each derivation $t : \Gamma \vdash A$ in \L\ is interpreted as a
natural transformation $\tmsem{t} : \cxtsem{\Gamma} \nattrans
\fmasem{A}$. The evaluation function is defined by recursion on
the argument $t$. The interesting cases are: $\II$ and $\otI$, where an extra
application of the monad unit $\eta$ is required; $\IE$ and $\otE$, which
involves both monoidal strengths $\lmst$ and $\rmst$, and the $\run$
function.
\begin{equation}\label{eq:eval}
  \begin{array}{lcl}
    \tmsem{\ax} \; a & \eqdf & a \\

    \tmsem{\olI\;t} \;\gamma & \eqdf & \lambda x.\,
    \tmsem{t}\;(\gamma,x) \\

    \tmsem{\loI\;t} \;\gamma & \eqdf & \lambda x.\,
    \tmsem{t}\;(x,\gamma) \\
    
    \tmsem{\II}\;\gamma & \eqdf & \eta\;\gamma \\

    \tmsem{\otI\;t\;u}\;(\gamma,\delta) & \eqdf &
    \eta\;(\tmsem{t}\;\gamma,\tmsem{u}\;\delta) \\

    \tmsem{\olE\;t\;u} \; (\gamma,\delta) & \eqdf &
    \tmsem{t}\;\gamma\;(\tmsem{u}\;\delta) \\

    \tmsem{\loE\;t\;u} \; (\gamma,\delta) & \eqdf &
    \tmsem{u}\;\delta\;(\tmsem{t}\;\gamma) \\

    \tmsem{\IE\;t\;u}\;(\delta_0,\gamma,\delta_1) & \eqdf & \run
    \; (T\;\tmsem{u}\;(\rmst\;(\lmst\;(\delta_0,\tmsem{t}\;\gamma),\delta_1))) \\
    
    \tmsem{\otE\;t\;u}\;(\delta_0,\gamma,\delta_1) & \eqdf & \run
    \; (T\;\tmsem{u}\;(\rmst\;(\lmst\;(\delta_0,\tmsem{t}\;\gamma),\delta_1)))
  \end{array}
\end{equation}
Pictorially, $\tmsem{\otE\;t\;u}$ is the following composite natural
transformation (remember that in our definitions, and in particular in the
definition of $\tmsem{\otE\;t\;u}$ above, applications of the
associativity natural isomorphism of $\otp$ are omitted):
\begin{equation*}
  \xymatrix@R=.1pc@C=3pc{
    \cxtsem{\Delta_0,\Gamma,\Delta_1}
    \ar[r]^-{\cong} &
    (\cxtsem{\Delta_0} \otp \cxtsem{\Gamma}) \otp \cxtsem{\Delta_1}
    \ar[r]^-{(\id \otp \tmsem{t}) \otp \id} &
    (\cxtsem{\Delta_0}\otp T\;(\fmasem{A} \otp \fmasem{B})) \otp \cxtsem{\Delta_1}
    \\ &
    \ar[r]^-{\lmst \otp \id} & 
    T\;(\cxtsem{\Delta_0}\otp(\fmasem{A} \otp \fmasem{B})) \otp \cxtsem{\Delta_1}
    \\ &
    \ar[r]^-{\rmst} &
    T\;((\cxtsem{\Delta_0}\otp(\fmasem{A} \otp \fmasem{B})) \otp \cxtsem{\Delta_1})
    \\ &
    \ar[r]^-{\cong} &
    T\;(\cxtsem{\Delta_0,A,B,\Delta_1})
    \\ &
    \ar[r]^-{T\;\tmsem{u}} &  T\;\fmasem{C} \ar[r]^{\run_C} & \fmasem{C}
    }
\end{equation*}

The evaluation function $\tmsem{-}$ is well-defined on
$\deq$-equivalence classes: given two derivations $t,u : \Gamma \vdash
A$, if $t \deq u$ then $\tmsem{t} = \tmsem{u}$. The proof of this fact
relies on $T$ being a strong monad and $\run$ being an algebra for the
monad $T$. For example $\run_A\;(\eta\;x) = x$, for all $x :
\fmasem{A}\;\Gamma$.

The interpretation of derivations can be readily extended to
environments $\envsem{\sigma} : \cxtsem{\Gamma} \nattrans
\cxtsem{\Delta}$, for each $\sigma : \Gamma \tr \Delta$:
  $\envsem{\nil} \;x\eqdf x$ and 
  $\envsem{(t,\sigma)}\;(\gamma,\gamma') \eqdf (\tmsem{t}\;\gamma,\envsem{\sigma}\;\gamma')$.
The evaluation function behaves well wrt. substitution: $\tsem{\sub\;\sigma\;t} = \tsem{t} \circ \envsem{\sigma}$, i.e. the cut rule is interpreted as function composition in the Kripke model.

\subsection{The Normalization Function}\label{sec:nbefun}

The last phase of the NbE procedure is the extraction of normal forms
from elements of the Kripke model. Concretely, this corresponds to the
construction of a \emph{reification} function $\reify_A : \fmasem{A}
\ \nattrans \ - \up A$. In order to deal with the cases of
mixed-variance connectives ${\ol}$ and ${\lo}$, it is necessary to
simultaneously define a \emph{reflection} procedure $\reflect_A : -
\dn A \ \nattrans \ \fmasem{A}$, embedding neutrals in the presheaf
model.  This is the crucial point were the interpretation of the unit
$\fmasem{\I} \eqdf T\;\Ip$ and the tensor product $\fmasem{A \ot B}
\eqdf T\;(\fmasem{A} \otp \fmasem{B})$ in (\ref{eq:fma}) works, while
na{\"i}ve interpretations $\fmasem{\I} \eqdf \Ip$ and $\fmasem{A \ot
  B} \eqdf \fmasem{A} \otp \fmasem{B}$ without the application of the
monad $T$ would fail. With the latter interpretation,
$\reflect_{\I}\;t$ would be required to have type $\Ip \;\Gamma$,
which in turn will force us to show that the context $\Gamma$ is
empty. But this is generally false, e.g. when $t$ is of the form $\ax:
\I \dn \I$. Analogously, $\reflect_{A \ot B}\;t$ would be required to
have type $(\fmasem{A} \otp \fmasem{B})\;\Gamma$, which in turn will
force us to split the context $\Gamma = \Gamma_0,\Gamma_1$ and provide
elements of type $\fmasem{A}\;\Gamma_0$ and
$\fmasem{B}\;\Gamma_1$. But this split is generally impossible to
achieve, e.g. for $t$ of the form $\ax : A \ot B \dn A \ot B$. This
problematic splitting in the definition of reflection $\reflect_A$ is
avoided by the use of the monad $T$, and the cases of $A$ being
unit or tensor are dealt via the application of the constructors
$\IET$ and $\otET$. The employment of a strong monad seems to be a general pattern in NbE for calculi including positive logical connectives, e.g. consider falsity and disjunction in
intuitionistic propositional logic~\cite{AU:nore,AS:norecp}. 
\begin{equation}\label{eq:reify}
  \begin{array}{lcl}

    \reify_p \;t & \eqdf & t \\

    \reify_{B \ol A} \;t & \eqdf & \olI \;(\reify_B (t\;(\reflect_A
    \ax)))\\

    \reify_{A \lo B} \;t & \eqdf & \loI \;(\reify_B (t\;(\reflect_A
    \ax)))\\
    
    \reify_{\I} \;t & \eqdf & \runup\;(T\;(\lambda \,\refl.\,\II)\;t) \\

    \reify_{A \ot B} \;t & \eqdf & \runup\;(T\;(\lambda (x,y).\,
    \otI\;(\reify_A x)\;(\reify_B y))\;t)    
  \end{array}
  \quad
  \begin{array}{lcl}
    \reflect_p  t & \eqdf & \sw t \\

    \reflect_{B \ol A} t & \eqdf & \lambda x.\, \reflect_B (\olE
    \;t\;(\reify_A x))\\

    \reflect_{A \lo B} t & \eqdf & \lambda x.\, \reflect_B
    (\loE\;(\reify_A x)\;t) \\

    \reflect_{\I} t & \eqdf & \IET\;t\;(\eta\;\refl) \\

    \reflect_{A \ot B} t & \eqdf & \otET\;t\;(\eta\;(\reflect_A
    \ax,\reflect_B \ax))    
  \end{array}  
\end{equation}
The reflection function $\reflect_A$ can also be used for the
definition of an element $\fresh_\Gamma : \cxtsem{\Gamma}\;\Gamma$,
for each context $\Gamma$: 
%
  $\fresh_{\nil} \eqdf \refl$ and 
%
    $\fresh_{A,\Gamma} \eqdf (\reflect_A \ax,\fresh_\Gamma)$.
When $\Gamma = \nil$, the element $\fresh_{\nil}$ has type
$\cxtsem{\nil}\;\nil$, which reduces to $\nil = \nil$, and $\refl$ is
the proof of reflexivity of equality.

The normalization function $\nbe : \Gamma \vdash A \to \Gamma \up A$
is then definable as the reification of the evaluation of a
derivation $t : \Gamma \vdash A$ in the Kripke model:
\begin{equation*}
  \begin{array}{lcl}
    \nbe\;t & \eqdf & \reify_A (\tmsem{t}\;\fresh_\Gamma)    
  \end{array}
\end{equation*}
Here we consider the interpretation $\tmsem{t} :
\cxtsem{\Gamma}\;\Gamma \to \fmasem{A} \;\Gamma$, which can be applied
to $\fresh_\Gamma : \cxtsem{\Gamma}\;\Gamma$.
Since $\tmsem{t} = \tmsem{u}$ for all $t \deq u$, then the
function $\nbe$ sends $\deq$-related derivations in \L\ to equal
derivations in \Lup, i.e. $\nbe\;t = \nbe\;u$ whenever $t\deq u$.

\begin{example}
Let $p,q,r$ be atomic formulae and let $t : p \ot q , r \vdash
p \ot (q \ot r)$ be the following derivation:
\[
\infer[\otE]{p \ot q , r \vdash p \ot (q \ot r)}{
  \infer[\ax]{p\ot q \vdash p\ot q}{}
  &
  \infer[\olE]{p , q , r \vdash p \ot (q \ot r)}{
    \infer[\otE]{p,q \vdash (p \ot (q \ot r)) \ol r}{
      \infer[\otI]{p,q \vdash p \ot q}{
        \infer[\ax]{p \vdash p}{}
        &
        \infer[\ax]{q \vdash q}{}
      }
      &
      \infer[\olI]{p,q \vdash (p \ot (q \ot r)) \ol r}{
        \infer[\otI]{p,q,r \vdash p \ot (q \ot r)}{
          \infer[\ax]{p \vdash p}{}
          &
          \infer[\otI]{q,r \vdash q \ot r}{
            \infer[\ax]{q \vdash q}{} &
            \infer[\ax]{r \vdash r}{}
          }
        }
      }
    }
    &
    \infer[\ax]{r \vdash r}{}
  }
}
\]
It is possible to check that the evaluation $\tmsem{t}$, when applied
to the element $\fresh_{p \ot q , r} : \cxtsem{p \ot q , r}\;(p \ot q
, r)$, returns the following element of $T\;(\fmasem{p} \otp (T\;(\fmasem{q} \otp \fmasem{r}))\;(p
\ot q , r)$:
\begin{equation}\label{eq:ex}
\tmsem{t}\;(\fresh_{p \ot q , r}) \; = \; \;
\proofbox{
\infer[\otET]{T\;(\fmasem{p} \otp (T\;(\fmasem{q} \otp \fmasem{r})))\;(p \ot q
  , r)}{
  \infer[\ax]{p \ot q \dn p \ot q}{}
  &
  \infer[\eta]{T\;(\fmasem{p} \otp (T\;(\fmasem{q} \otp \fmasem{r})))\;(p, q , r)}{
    \infer[\sw]{p \up p}{\infer[\ax]{p \dn p}{}}
    &
    \infer[\eta]{T\;(\fmasem{q} \otp \fmasem{r})\;(q , r)}{
      \infer[\sw]{q \up q}{\infer[\ax]{q \dn q}{}}
      &
      \infer[\sw]{r \up r}{\infer[\ax]{r \dn r}{}}      
    }
  }
}
}
\end{equation}
(In the applications of the monad unit $\eta$ above, the definition of
$\otp$ is automatically unfolded). The application of the function
$\nbe$ to $t$, corresponding to the application of the reification map
$\reify_{p \ot (q \ot r)}$ to the element in (\ref{eq:ex}), returns
the following $\beta\eta$-long normal form in \Lup:
\[
\nbe\;t  \; = \; \;
\proofbox{
  \infer[\otE]{p \ot q , r \up p \ot (q \ot r)}{
    \infer[\ax]{p \ot q \dn p \ot q}{}
    &
    \infer[\otI]{p , q , r \up p \ot (q \ot r)}{
      \infer[\sw]{p \up p}{\infer[\ax]{p \dn p}{}}
      &
      \infer[\otI]{q , r \up q \ot r}{
        \infer[\sw]{q \up q}{\infer[\ax]{q \dn q}{}}
        &
        \infer[\sw]{r \up r}{\infer[\ax]{r \dn r}{}}      
      }
    }
  }
}
\]
\end{example}

\section{Correctness of NbE}\label{sec:corr}

So far we have constructed an effective normalization procedure
sending $\beta\eta$-related derivations in \L\ to the same normal form
in \Lup. It remains to show that this procedure is \emph{correct},
which is to say that $\nbe$ is a bijective function with $\embup$ as
its inverse, up to the equivalence of derivations $\deq$. This in turn implies that each derivation in \L\ is
$\beta\eta$-equivalent to (the embedding of) its normal form, and each
derivation in \Lup\ is the unique representative of a
$\deq$-equivalence class in \L.

The latter property of the normalization function, corresponding to
the surjectivity of $\nbe$, is straightforward. Formally, it is
possible to prove the following statements by structural induction
on the input normal forms and neutrals.
\begin{description}
  \item[(surjectivity of $\nbe$)] For any neutral $n : \Gamma \dn A$, we have 
    $\reflect_A n = \tmsem{\embdn\;n}\;\fresh_\Gamma$. \\For any normal form $n : \Gamma \up
    A$, we have $n = \nbe\;(\embup n)$. 
\end{description}

The injectivity of the normalization function $\nbe$ can be proved
employing a \emph{logical relation} $\approx_A$, which relates each
derivation of a sequent $\Gamma \vdash A$ in \L\ with the corresponding denoting values
of $\fmasem{A}\;\Gamma$ in the Kripke model. It is mutually-inductively defined
together with another relation $\approx^\ot$, whose definition is given below. The generating inference rules of $\approx_A$ are:
\begin{equation}\label{eq:logrel}
  \def\arraystretch{2.2}
  \begin{array}{c}
  \infer{t \approx_p n}{t \deq \embup\, n}
  \qquad
  \infer{t \approx_{\I} v}{t \, (\overline{T}{\approx}^\I)\, v}
  \qquad
  \infer{t \approx_{A \ot B} v}{t \, (\overline{T}{\approx}^\ot)\, v}
  \\
  \infer{t \approx_{B \ol A} f}{\{\Delta : \Cxt \} \; (u : \Delta \vdash A) \;(a : \fmasem{A}\;\Delta) \to u \approx_A a \to \olE\; t\; u \approx_B f \;a}
  \\
  \infer{t \approx_{A \lo B} f}{\{\Delta : \Cxt \} \; (u : \Delta \vdash A) \;(a : \fmasem{A}\;\Delta) \to u \approx_A a \to \loE\; u\; t \approx_B f \;a}
  \end{array}
\end{equation}
The definition of the logical relation is very similar to the one used
to prove correctness of normalization for (the implicational fragment
of) ordered linear logic~\cite{Pol:ordlla}, but the cases of the
positive connectives $\I$ and $\ot$ require additional care because of
the presence of the monad $T$ in the interpretation of formulae. To
this end, the rules in (\ref{eq:logrel}) make use of the following
auxiliary notions:
\begin{itemize}
\item Given a relation $R \subseteq (\Gamma \vdash C) \times
  (P\;\Gamma)$ between syntactic derivations and semantic values, the
  \emph{lifting} of $R$ to the monad $T$ is a relation $\overline{T}R
  \subseteq (\Gamma \vdash C) \times (T\;P\;\Gamma)$,
  whose validity is specified inductively by the following rules:
  \begin{equation*}
    \def\arraystretch{2.2}
    \begin{array}{c}
    \infer{t\,(\overline{T}R)\,(\eta\;x)}{t\,R\,x}
    \qquad
    \infer{t\,(\overline{T}R)\,\IET\;u\;v}{
      t' : \Delta_0,\Delta_1 \vdash C
      &
      t \deq \IE \;(\embdn u)\;t'
      &
      t' \,(\overline{T}R)\,v
    } \\
     \infer{t\,(\overline{T}R)\,\otET\;u\;v}{
      t' : \Delta_0,A,B,\Delta_1 \vdash C
      &
      t \deq \otE \;(\embdn u)\;t'
      &
      t' \,(\overline{T}R)\,v
    }
    \end{array}
  \end{equation*}
  The second rule should be read as: a
  derivation $t : \Delta_0, \Gamma, \Delta_1 \vdash C$ is $(\overline{T}R)$-related to $\IET\;u\;v$, for some neutral $u : \Gamma \dn \I$ and monadic value $v : T\;P\;(\Delta_0,\Delta_1)$, if and only if $t$ is $\beta\eta$-equivalent to a unit-elimination $\IE\;(\embdn u)\;t'$, for some derivation $t' : \Delta_0,\Delta_1 \vdash C$, such that $t' \,(\overline{T}R)\,v$. The third rule should be read analogously.
\item The relation $\approx^\I \subseteq (\Gamma \vdash I) \times
  (\Ip\;\Gamma)$ is defined as follows: $t \approx^\I v$ if and only if
  $t \ceq \II$ (notice that $v : \Ip \;\Gamma$ forces $\Gamma$ to be
  empty).
\item The relation $\approx^\ot \subseteq (\Gamma \vdash A \ot B)
  \times ((\fmasem{A} \otp \fmasem{B})\,\Gamma)$ is defined as follows: $t
  \approx^\ot (a,b)$ if and only if $u \,\approx_A \,a$ and $v \,\approx_B \,b$ for some derivations $u : \Gamma_0 \vdash A$ and $v : \Gamma_1 \vdash B$ such that $t \deq \otI\;u\;v$ (notice that by assumption $\Gamma$ is of the form $\Gamma_0,\Gamma_1$, so that $a : \fmasem{A}\;\Gamma_0$ and $b : \fmasem{B}\;\Gamma_1$).
 In particular, the relation $\approx^\ot$ is defined simultaneously with $\approx$.
\end{itemize}

The logical relation $\approx$ is invariant under
$\beta\eta$-conversion, formally:
\begin{description}
  \item[(fundamental theorem of logical relations)] For
any derivations $t,u : \Gamma \vdash A$ and value $v : \fmasem{A}\;\Gamma$, if $t \deq
u$ and $u \approx_A v$, then also $t \approx_A v$.
\end{description}
This fact can be proved by induction on the proof of $u \approx_A v$,
together with similar closure properties of the relation lifting
$\overline{T}$ and the auxiliary relations $\approx^\I$ and
$\approx^\ot$.

It is possible to define a relation $\approx_\Gamma^{\tr}\subseteq (\Gamma
\tr \Delta) \times (\envsem{\Gamma}\;\Delta)$ which naturally extends
the logical relation $\approx$ to environments and their
interpretations. Using the fundamental theorem, one can show that the
syntactic notion of substitution given by the admissible $\cut$ rule
is $\approx$-related to the semantic notion of substitution performed
during the interpretation of a formula in the Kripke model. Formally:
\begin{description}
  \item[(correctness of evaluation)] Given a derivation $t : \Gamma
    \vdash A$, a syntactic environment $\sigma : \Gamma \tr \Delta$
    and a semantic environment $\gamma : \envsem{\Gamma}\;\Delta$ such
    that $\sigma \, \approx_\Gamma^\tr \, \gamma$, we have $\cut\;\sigma\;t
    \approx_A \tmsem{t}\;\gamma$.
\end{description}

From the fundamental theorem it follows that the
reification and reflection functions behave correctly:
\begin{description}
  \item[(correctness of reification)] For any derivation $t : \Gamma \vdash C$ and semantic value $v : \fmasem{A}\;\Gamma$, if $t \approx v$ then $t \deq \embup\;(\reify_A v)$. This means that whenever  $v$ denotes $t$, then the reification of $v$ (seen as a derivation in \L\ via $\embup$) is correctly $\beta\eta$-related to $t$.
\item[(correctness of reflection)] For any neutral $n : \Gamma \dn A$, we have $\embdn\;n \,\approx\; \reflect_A n$. This means that the reflected neutral $n$ correctly denotes $n$ (seen as a derivation in \L\ via $\embdn$).
\end{description}

The desired correctness property of $\nbe$ can therefore be formally stated as:
\begin{description}
  \item[(injectivity of $\nbe$)] For all derivations $t : \Gamma \vdash A$, we have $t \deq \embup \;(\nbe\;t)$.
\end{description}
This can be proved by connecting all the correctness results obtained so far:
\[
\begin{array}{rcl@{\hspace{1.9pt}}r}
  t \deq \embup \;(\nbe\;t) & \text{iff} & t \deq \embup (\reify_A (\tmsem{t}\;\fresh_\Gamma))
  & (\text{unfolding the definition of $\nbe$}) \\
  & \text{true if} & t \approx_A \tmsem{t}\;\fresh_\Gamma
  & (\text{correctness of reification}) \\
  & \text{true if} & \cut\;\ids_\Gamma\;t \approx_A \tmsem{t}\;\fresh_\Gamma
  & (\text{fundamental theorem applied to $t \sim \cut\;\ids_\Gamma\;t$}) \\
  & \text{true if} & \ids_\Gamma \approx_\Gamma^\tr \fresh_\Gamma
  & (\text{correctness of evaluation)}
\end{array}
\]
and $\ids_\Gamma \approx_\Gamma^\tr \fresh_\Gamma$ is a simple consequence of the
correctness of reflection.

\section{NbE for Fragments of Intuitionistic Linear Logic}\label{sec:milldill}

The NbE strategy described so far for the Lambek calculus works also for other related substructural systems, such as multiplicative intuitionistic linear logic (MILL) and dual intuitionistic linear logic (DILL).

\subsection{Multiplicative Intuitionistic Linear Logic}

A natural deduction calculus for MILL~\cite{BBdePH:tercil} is obtained from \L\ by adding an explicit exchange rule allowing to swap the positions of any two formulae in the context. Alternatively, and this is the modification we choose, one could also change the definition of a sequent $\Gamma \vdash A$ by requiring the context $\Gamma$ to be a finite multiset of formulae instead of an ordered list. In particular, the comma notation in $\Gamma,\Delta$ now stands for multiset concatenation. The change of nature of antecedents makes the presence of two implications $\lo$ and $\ol$ redundant, since they become equivalent connectives. The two ordered implications are then replaced by a unique linear implication $\multimap$, and only one occurrence of the two implication introduction and elimination rules is preserved, similarly for the equations in Figure~\ref{eq:eqs1}. All the other rules and equations are the same. Similar modifications occur in the calculus of normal forms \Lup.

The construction of the Kripke model and the implementation of the NbE procedure for \L\ can then be directly adapted to the case of MILL. The main difference is that the presheaf model $\Set^\Cxt$ is now \emph{symmetric} monoidal closed, which means that there is a natural isomorphism $P \otp Q \cong Q \otp P$.

\subsection{Dual Intuitionistic Linear Logic}

Our NbE strategy extends with some modifications also to the case of dual intuitionistic linear logic (DILL)~\cite{BP:duaill}, which is a particular presentation of MILL with the exponential modality !. Its peculiarity comes from the design of sequents, which are triples of the form $\Gamma;\Delta \vdash A$, where $\Gamma$ is a list of formulae (the \emph{intuitionistic context}) and $\Delta$ is a finite multiset of formulae (the \emph{linear context}). The rules of DILL are listed in Figure~\ref{eq:dill} (the $\beta\eta$-equivalence on DILL derivations is omitted from the paper). The rules for $\beta\eta$-long normal forms and neutrals are presented in Figure~\ref{eq:dillup}.

The presence of the intuitionistic context makes the construction of the Kripke model slightly more involved. Let $\LCxt$ be the set of linear contexts and $\ICxt$ be the category with intuitionistic contexts as objects and \emph{renamings} as maps, i.e. injective functions $f : \Gamma \to \Gamma'$ sending each occurrence of a formula in $\Gamma$ to a distinct occurrence of a formula in $\Gamma'$. The Kripke model for DILL is the (covariant) presheaf category $\Set^{\ICxt \times \LCxt}$. This category has again a (symmetric) monoidal closed structure, which is defined analogously to (\ref{eq:monbi}). For example, the (action on objects of the) semantics tensor product is $(P \otp Q) \; (\Gamma;\Delta) \eqdf \{\Delta_0,\Delta_1 : \LCxt\} \times \{ \Delta = \Delta_0 , \Delta_1 \} \times P\;(\Gamma;\Delta_0) \times Q\;(\Gamma;\Delta_1)$.
This category has an exponential modality $\bp$ whose action on objects is given by
\[
\bp \;P \;(\Gamma;\Delta) \eqdf (\Delta = \nil) \times P\;(\Gamma;~)
\]
The definition of the monad $T$ on $\Set^{\ICxt \times \LCxt}$ is analogous to the one in (\ref{eq:monad}) with the following addition
\[
\infer[\bET]{T\;P\;(\Gamma;\Delta,\Delta')}{
    \Gamma;\Delta \dn !A &
    T\;P\;(\Gamma,A;\Delta')
  }
\]
and the interpretation $\fmasem{!A}$ is given by $T\;(\bp\;\fmasem{A})$.
The interpretation of formulae extends to contexts $\cxtsem{\Gamma;\Delta}(\Gamma';\Delta') \eqdf \cxtsem{\Gamma}^\times\Gamma' \, \times \, \cxtsem{\Delta}^\ot(\Gamma';\Delta')$, where the interpretation of linear contexts $\cxtsem{\Delta}^\ot$ is as before (i.e. defined using tensor $\otp$) and the interpretation of intuitionistic contexts is defined using the Cartesian product of the Kripke model, i.e.  $\cxtsem{A_1,\dots,A_n}^\times\;\Gamma' \eqdf\fmasem{A_1}\;(\Gamma';~) \times \dots \times \fmasem{A_n}\;(\Gamma';~)$.
\begin{figure}[t!]
\begin{equation*}
  \def\arraystretch{1.4}
  \begin{array}{c}
  \infer[\loi]{\Gamma;\Delta \vdash A \multimap B}{
    \Gamma;\Delta, A \vdash B
  }  
  \qquad
  \infer[\loe]{\Gamma;\Delta,\Delta' \vdash B}{
    \Gamma;\Delta \vdash A \multimap B & \Gamma;\Delta' \vdash A
  }
  \qquad
    \infer[\bI]{\Gamma;~ \vdash !A}{
    \Gamma;~ \vdash A
  }
  \qquad
  \infer[\bE]{\Gamma;\Delta,\Delta' \vdash C}{
    \Gamma;\Delta \vdash !A & \Gamma,A;\Delta' \vdash C
  }
  \\
    \infer[\ax_{Lin}]{\Gamma;A \vdash A}{}
    \qquad
    \infer[\ax_{Int}]{\Gamma_0,A,\Gamma_1;~ \vdash A}{}
    \qquad
  \infer[\II]{\Gamma; ~ \vdash \I}{}
  \qquad
  \infer[\IE]{\Gamma;\Delta , \Delta' \vdash C}{
    \Gamma;\Delta \vdash \I & \Gamma;\Delta' \vdash C
  }
  \\
  \infer[\otI]{\Gamma;\Delta,\Delta' \vdash A \ot B}{
    \Gamma;\Delta \vdash A & \Gamma;\Delta' \vdash B
  } 
  \qquad
  \infer[\otE]{\Gamma;\Delta , \Delta' \vdash C}{
    \Gamma;\Delta \vdash A \ot B & \Gamma;\Delta' , A,B \vdash C
  }
  \end{array}
\end{equation*}
\caption{Inference rules of DILL}
\label{eq:dill}
\end{figure}
\begin{figure}[t!]
\begin{equation*}
  \def\arraystretch{1.4}
  \begin{array}{c}
  \infer[\loi]{\Gamma;\Delta \up A \multimap B}{
    \Gamma;\Delta, A \up B
  }
  \qquad
  \infer[\loe]{\Gamma;\Delta,\Delta' \dn B}{
    \Gamma;\Delta \dn A \multimap B & \Gamma;\Delta' \up A
  }
  \qquad
    \infer[\bI]{\Gamma;~ \up !A}{
    \Gamma;~ \up A
  }
  \qquad
  \infer[\bE]{\Gamma;\Delta,\Delta' \up C}{
    \Gamma;\Delta \dn !A & \Gamma,A;\Delta' \up C
  }  
    \\
    \infer[\ax_{Lin}]{\Gamma;A \dn A}{}
    \qquad
    \infer[\ax_{Int}]{\Gamma,A,\Gamma';~ \dn A}{}
    \qquad
      \infer[\II]{\Gamma;~ \up \I}{
  }
  \qquad
  \infer[\IE]{\Gamma;\Delta, \Delta' \up \gamma^+}{
    \Gamma;\Delta \dn \I & \Gamma;\Delta' \up \gamma^+
  }
  \\
    \infer[\otI]{\Gamma;\Delta,\Delta' \up A \ot B}{
    \Gamma;\Delta \up A & \Gamma;\Delta' \up B
  }
  \qquad
  \infer[\otE]{\Gamma;\Delta , \Delta' \up \gamma^+}{
    \Gamma;\Delta \dn A \ot B & \Gamma;\Delta' , A,B \up \gamma^+
  }
  \qquad
    \infer[\sw]{\Gamma;\Delta \up p}{
    \Gamma;\Delta \dn p}
\end{array}
\end{equation*}
\caption{Normal forms and neutrals of DILL}
\label{eq:dillup}
\end{figure}

The definition of evaluation $\tmsem{t} : \cxtsem{\Gamma;\Delta} \nattrans \fmasem{A}$ in (\ref{eq:eval}) needs to take into account the three new rules
\begin{equation*}
\begin{array}{lcl}
\tmsem{\ax_{Int}} \; (\gamma,a,\gamma') & \eqdf & a \\
\tmsem{\bI\;t}\;(\gamma,\delta) & \eqdf &
\eta\;(\tmsem{t}\;(\gamma,\delta)) \\
\tmsem{\bE\;t\;u}\;(\gamma,\delta_0,\delta_1) & \eqdf & \run
\; (T\;\tmsem{u}\;(\rmst\;(\lmst\;(\gamma,\tmsem{t}\;(\gamma,\delta_0)),\delta_1)))
\end{array}
\end{equation*}
where the function $\run$ is as in (\ref{eq:run}), with the extra case of $\bET$ dealt similarly to $\IET$ and $\otET$. The left strength  $\lmst$ (and the equivalent right strength $\rmst$) is defined as in (\ref{eq:lmst}), but the case of $\bET$ requires some extra care:
$\lmst\;(x,\bET\;t\;y) \eqdf \bET\;t\;(\lmst\;(P\;\iota\;x,y))$, where $P\;\iota : P \;(\Gamma;\Delta) \to P\;(\Gamma,A;\Delta)$ is the action of the presheaf $P$ on the renaming $\iota : \Gamma \to \Gamma,A$, which is the inclusion of $\Gamma$ into $\Gamma,A$.
Reflection $\reflect_A$ and reification $\reify_A$ in (\ref{eq:reify}) also need to cover the extra case of the modality !:
\[
  \begin{array}{lcl}
    \reify_{!A} \;t & \eqdf & \runup\;(T\;(\lambda (\refl,y).\,
    \bI\;(\reify_A y))\;t)
  \end{array}
  \qquad\quad
  \begin{array}{lcl}
    \reflect_{!A} t & \eqdf & \bET\;t\;(\eta\;(\refl,\reflect_A
    \ax_{Int}))
  \end{array}
\]
This implies that the $\nbe$ function can also be defined on DILL derivations. Similarly to the case of \L, the correctness of $\nbe$ can be proved via logical relations. A thorough description of the NbE procedure for DILL, together with a proof of its correctness, will appear
in a future extended version of this paper.

\section{Conclusions}
\label{sec:concl}

We have studied normalization by evaluation for the Lambek calculus and other related substructural logics such as MILL and DILL. The presence of positive connectives, i.e. the multiplicative unit $\I$ and tensor $\ot$ (and exponential $!$ in DILL), requires a modification of the NbE technique for typed $\lambda$-calculus employing  a strong monad on the Kripke model, a situation sharing many similarities with the extension of typed $\lambda$-calculus with empty and sum types~\cite{AU:nore,AS:norecp}.

The normalization algorithm developed in this paper works for ``weak''
unit and tensor, in the sense that the equational theory
$\deq$ does not satisfy equations of the form (\ref{eq:missing}). In
other words, \Lup\ is not a presentation of the free monoidal biclosed
category on the set of atomic formulae $\At$. The extension of
NbE to ``strong'' unit and tensor is left for future work. We should
take inspiration from the methodologies adopted in the NbE literature
to deal with strong sums in typed
$\lambda$-calculus~\cite{ADHS:noretl,BCF:extntd}, which require a
modification of the Kripke semantics using sheaves on a particularly
chosen site instead of presheaves. The normal forms for the stronger
equational theory would also serve as the target calculus of the
normalization procedure of Amblard and Retor{\'e}~\cite{AR:norspl}

This paper should be considered as a starting point in understanding the NbE methodology in the realm of substructural logics. We are especially interested in the connection between Kripke logical relations for linear logic~\cite{Has:logpil,HS:gluoml} and categorical reformulations of normalization by evaluation~\cite{AHS:catrrf,Fio:semane}.

\paragraph{Acknowledgments}
The author was supported by the ESF funded Estonian IT Academy research measure (project 2014-2020.4.05.19-0001) and the Estonian Research Council grants PSG659 and PSG749. Participation in the conference was supported by the EU COST
action CA19135 (CERCIRAS).

\bibliographystyle{eptcs}
\bibliography{biblio}

\begin{thebibliography}{10}
\providecommand{\bibitemdeclare}[2]{}
\providecommand{\surnamestart}{}
\providecommand{\surnameend}{}
\providecommand{\urlprefix}{Available at }
\providecommand{\url}[1]{\texttt{#1}}
\providecommand{\href}[2]{\texttt{#2}}
\providecommand{\urlalt}[2]{\href{#1}{#2}}
\providecommand{\doi}[1]{doi:\urlalt{http://dx.doi.org/#1}{#1}}
\providecommand{\eprint}[1]{arXiv:\urlalt{https://arxiv.org/abs/#1}{#1}}
\providecommand{\bibinfo}[2]{#2}

\bibitemdeclare{inproceedings}{AS:norecp}
\bibitem{AS:norecp}
\bibinfo{author}{Andreas \surnamestart Abel\surnameend} \&
  \bibinfo{author}{Christian \surnamestart Sattler\surnameend}
  (\bibinfo{year}{2019}): \emph{\bibinfo{title}{Normalization by Evaluation for
  Call-by-Push-Value and Polarized Lambda Calculus}}.
\newblock In \bibinfo{editor}{Ekaterina \surnamestart
  Komendantskaya\surnameend}, editor: {\sl \bibinfo{booktitle}{Proceedings of
  the 21st International Symp. on Principles and Practice of Programming
  Languages, PPDP 2019}}, \bibinfo{publisher}{{ACM}}, pp.
  \bibinfo{pages}{3:1--3:12}, \doi{10.1145/3354166.3354168}.

\bibitemdeclare{article}{Abr:noncil}
\bibitem{Abr:noncil}
\bibinfo{author}{Vito~M. \surnamestart Abrusci\surnameend}
  (\bibinfo{year}{1990}): \emph{\bibinfo{title}{Non‐Commutative
  Intuitionistic Linear Logic}}.
\newblock {\sl \bibinfo{journal}{Mathematical Logic Quarterly}}
  \bibinfo{volume}{36}(\bibinfo{number}{4}), pp. \bibinfo{pages}{297--318},
  \doi{10.1002/malq.19900360405}.

\bibitemdeclare{inproceedings}{ADHS:noretl}
\bibitem{ADHS:noretl}
\bibinfo{author}{Thorsten \surnamestart Altenkirch\surnameend},
  \bibinfo{author}{Peter \surnamestart Dybjer\surnameend},
  \bibinfo{author}{Martin \surnamestart Hofmann\surnameend} \&
  \bibinfo{author}{Philip~J. \surnamestart Scott\surnameend}
  (\bibinfo{year}{2001}): \emph{\bibinfo{title}{Normalization by Evaluation for
  Typed Lambda Calculus with Coproducts}}.
\newblock In: {\sl \bibinfo{booktitle}{Proceedings of the 16th Annual {IEEE}
  Symp. on Logic in Computer Science, LICS 2001}}, \bibinfo{publisher}{{IEEE}
  Computer Society}, pp. \bibinfo{pages}{303--310},
  \doi{10.1109/LICS.2001.932506}.

\bibitemdeclare{inproceedings}{AHS:catrrf}
\bibitem{AHS:catrrf}
\bibinfo{author}{Thorsten \surnamestart Altenkirch\surnameend},
  \bibinfo{author}{Martin \surnamestart Hofmann\surnameend} \&
  \bibinfo{author}{Thomas \surnamestart Streicher\surnameend}
  (\bibinfo{year}{1995}): \emph{\bibinfo{title}{Categorical Reconstruction of a
  Reduction Free Normalization Proof}}.
\newblock In \bibinfo{editor}{David~H. \surnamestart Pitt\surnameend},
  \bibinfo{editor}{David~E. \surnamestart Rydeheard\surnameend} \&
  \bibinfo{editor}{Peter~T. \surnamestart Johnstone\surnameend}, editors: {\sl
  \bibinfo{booktitle}{Proceedings of the 6th International Conference on
  Category Theory and Computer Science, {CTCS} 1995}}, {\sl
  \bibinfo{series}{Lecture Notes in Computer Science}} \bibinfo{volume}{953},
  \bibinfo{publisher}{Springer}, pp. \bibinfo{pages}{182--199},
  \doi{10.1007/3-540-60164-3\_27}.

\bibitemdeclare{inproceedings}{AU:nore}
\bibitem{AU:nore}
\bibinfo{author}{Thorsten \surnamestart Altenkirch\surnameend} \&
  \bibinfo{author}{Tarmo \surnamestart Uustalu\surnameend}
  (\bibinfo{year}{2004}): \emph{\bibinfo{title}{Normalization by Evaluation for
  $\lambda$\(^{\mbox{${\to}2$}}\)}}.
\newblock In \bibinfo{editor}{Yukiyoshi \surnamestart Kameyama\surnameend} \&
  \bibinfo{editor}{Peter~J. \surnamestart Stuckey\surnameend}, editors: {\sl
  \bibinfo{booktitle}{Proceedings of the 7th International Symp. on Functional
  and Logic Programming, {FLOPS} 2004}}, {\sl \bibinfo{series}{Lecture Notes in
  Computer Science}} \bibinfo{volume}{2998}, \bibinfo{publisher}{Springer}, pp.
  \bibinfo{pages}{260--275}, \doi{10.1007/978-3-540-24754-8\_19}.

\bibitemdeclare{article}{AR:norspl}
\bibitem{AR:norspl}
\bibinfo{author}{Maxime \surnamestart Amblard\surnameend} \&
  \bibinfo{author}{Christian \surnamestart Retor{\'{e}}\surnameend}
  (\bibinfo{year}{2014}): \emph{\bibinfo{title}{Partially Commutative Linear
  Logic and Lambek Caculus with Product: Natural Deduction, Normalisation,
  Subformula Property}}.
\newblock {\sl \bibinfo{journal}{{FLAP}}}
  \bibinfo{volume}{1}(\bibinfo{number}{1}), pp. \bibinfo{pages}{53--94}.
\newblock
  \urlprefix\url{http://www.collegepublications.co.uk/downloads/ifcolog00001.pdf}.

\bibitemdeclare{inproceedings}{BCF:extntd}
\bibitem{BCF:extntd}
\bibinfo{author}{Vincent \surnamestart Balat\surnameend},
  \bibinfo{author}{Roberto~Di \surnamestart Cosmo\surnameend} \&
  \bibinfo{author}{Marcelo~P. \surnamestart Fiore\surnameend}
  (\bibinfo{year}{2004}): \emph{\bibinfo{title}{Extensional Normalisation and
  Type-Directed Partial Evaluation for Typed Lambda Calculus with Sums}}.
\newblock In \bibinfo{editor}{Neil~D. \surnamestart Jones\surnameend} \&
  \bibinfo{editor}{Xavier \surnamestart Leroy\surnameend}, editors: {\sl
  \bibinfo{booktitle}{Proceedings of the 31st {ACM} {SIGPLAN-SIGACT} Symposium
  on Principles of Programming Languages, {POPL} 2004}},
  \bibinfo{publisher}{{ACM}}, pp. \bibinfo{pages}{64--76},
  \doi{10.1145/964001.964007}.

\bibitemdeclare{phdthesis}{BP:duaill}
\bibitem{BP:duaill}
\bibinfo{author}{Andrew~G. \surnamestart Barber\surnameend}
  (\bibinfo{year}{1997}): \emph{\bibinfo{title}{Linear Type Theories, Semantics
  and Action Calculi}}.
\newblock Ph.D. thesis, \bibinfo{school}{University of Edinburgh, {UK}}.
\newblock \urlprefix\url{http://hdl.handle.net/1842/392}.

\bibitemdeclare{inproceedings}{BBdePH:tercil}
\bibitem{BBdePH:tercil}
\bibinfo{author}{Nick \surnamestart Benton\surnameend},
  \bibinfo{author}{Gavin~M. \surnamestart Bierman\surnameend},
  \bibinfo{author}{Valeria \surnamestart de~Paiva\surnameend} \&
  \bibinfo{author}{Martin \surnamestart Hyland\surnameend}
  (\bibinfo{year}{1993}): \emph{\bibinfo{title}{A Term Calculus for
  Intuitionistic Linear Logic}}.
\newblock In \bibinfo{editor}{Marc \surnamestart Bezem\surnameend} \&
  \bibinfo{editor}{Jan~Friso \surnamestart Groote\surnameend}, editors: {\sl
  \bibinfo{booktitle}{Proceedings of the 1st International Conference on Typed
  Lambda Calculi and Applications, {TLCA} '93}}, {\sl \bibinfo{series}{Lecture
  Notes in Computer Science}} \bibinfo{volume}{664},
  \bibinfo{publisher}{Springer}, pp. \bibinfo{pages}{75--90},
  \doi{10.1007/BFb0037099}.

\bibitemdeclare{inproceedings}{BS:inveft}
\bibitem{BS:inveft}
\bibinfo{author}{Ulrich \surnamestart Berger\surnameend} \&
  \bibinfo{author}{Helmut \surnamestart Schwichtenberg\surnameend}
  (\bibinfo{year}{1991}): \emph{\bibinfo{title}{An Inverse of the Evaluation
  Functional for Typed lambda-calculus}}.
\newblock In: {\sl \bibinfo{booktitle}{Proceedings of the 6th Annual Symposium
  on Logic in Computer Science, {LICS} 1991}}, \bibinfo{publisher}{{IEEE}
  Computer Society}, pp. \bibinfo{pages}{203--211},
  \doi{10.1109/LICS.1991.151645}.

\bibitemdeclare{inproceedings}{BNS:focnd}
\bibitem{BNS:focnd}
\bibinfo{author}{Taus \surnamestart Brock{-}Nannestad\surnameend} \&
  \bibinfo{author}{Carsten \surnamestart Sch{\"{u}}rmann\surnameend}
  (\bibinfo{year}{2010}): \emph{\bibinfo{title}{Focused Natural Deduction}}.
\newblock In \bibinfo{editor}{Christian~G. \surnamestart
  Ferm{\"{u}}ller\surnameend} \& \bibinfo{editor}{Andrei \surnamestart
  Voronkov\surnameend}, editors: {\sl \bibinfo{booktitle}{Proceedings of the
  17th International Conference on Logic for Programming, Artificial
  Intelligence, and Reasoning, {LPAR} 2010}}, {\sl \bibinfo{series}{Lecture
  Notes in Computer Science}} \bibinfo{volume}{6397},
  \bibinfo{publisher}{Springer}, pp. \bibinfo{pages}{157--171},
  \doi{10.1007/978-3-642-16242-8\_12}.

\bibitemdeclare{inproceedings}{Fio:semane}
\bibitem{Fio:semane}
\bibinfo{author}{Marcelo~P. \surnamestart Fiore\surnameend}
  (\bibinfo{year}{2002}): \emph{\bibinfo{title}{Semantic Analysis of
  Normalisation by Evaluation for Typed Lambda Calculus}}.
\newblock In: {\sl \bibinfo{booktitle}{Proceedings of the 4th International
  {ACM} {SIGPLAN} Conference on Principles and Practice of Declarative
  Programming, {PPDP} 2002}}, \bibinfo{publisher}{{ACM}}, pp.
  \bibinfo{pages}{26--37}, \doi{10.1145/571157.571161}.

\bibitemdeclare{inproceedings}{Has:logpil}
\bibitem{Has:logpil}
\bibinfo{author}{Masahito \surnamestart Hasegawa\surnameend}
  (\bibinfo{year}{1999}): \emph{\bibinfo{title}{Logical Predicates for
  Intuitionistic Linear Type Theories}}.
\newblock In \bibinfo{editor}{Jean{-}Yves \surnamestart Girard\surnameend},
  editor: {\sl \bibinfo{booktitle}{Proceedings of the 4th International
  Conference on Typed Lambda Calculi and Applications, {TLCA} 1999}}, {\sl
  \bibinfo{series}{Lecture Notes in Computer Science}} \bibinfo{volume}{1581},
  \bibinfo{publisher}{Springer}, pp. \bibinfo{pages}{198--212},
  \doi{10.1007/3-540-48959-2\_15}.

\bibitemdeclare{inproceedings}{Hep:norftp}
\bibitem{Hep:norftp}
\bibinfo{author}{Mark \surnamestart Hepple\surnameend} (\bibinfo{year}{1990}):
  \emph{\bibinfo{title}{Normal Form Theorem Proving for the Lambek Calculus}}.
\newblock In: {\sl \bibinfo{booktitle}{Proceedings of the 13th International
  Conference on Computational Linguistics, {COLING} 1990}}, pp.
  \bibinfo{pages}{173--178}.
\newblock \urlprefix\url{https://aclanthology.org/C90-2030/}.

\bibitemdeclare{article}{HS:gluoml}
\bibitem{HS:gluoml}
\bibinfo{author}{Martin \surnamestart Hyland\surnameend} \&
  \bibinfo{author}{Andrea \surnamestart Schalk\surnameend}
  (\bibinfo{year}{2003}): \emph{\bibinfo{title}{Glueing and Orthogonality for
  Models of Linear Logic}}.
\newblock {\sl \bibinfo{journal}{Theorerical Computer Science}}
  \bibinfo{volume}{294}(\bibinfo{number}{1/2}), pp. \bibinfo{pages}{183--231},
  \doi{10.1016/S0304-3975(01)00241-9}.

\bibitemdeclare{inproceedings}{LR:pronlc}
\bibitem{LR:pronlc}
\bibinfo{author}{Francois \surnamestart Lamarche\surnameend} \&
  \bibinfo{author}{Christian \surnamestart Retor{\'e}\surnameend}
  (\bibinfo{year}{1996}): \emph{\bibinfo{title}{Proof Nets for the Lambek
  Calculus -- an Overview}}.
\newblock In: {\sl \bibinfo{booktitle}{Proceedings of the 3rd Roma Workshop on
  Proofs and Linguistic Categories 1998}}, pp. \bibinfo{pages}{241--262}.
\newblock \urlprefix\url{https://hal.archives-ouvertes.fr/inria-00098442}.

\bibitemdeclare{article}{Lam:matss}
\bibitem{Lam:matss}
\bibinfo{author}{Joachim \surnamestart Lambek\surnameend}
  (\bibinfo{year}{1958}): \emph{\bibinfo{title}{The Mathematics of Sentence
  Structure}}.
\newblock {\sl \bibinfo{journal}{The American Mathematical Monthly}}
  \bibinfo{volume}{65}(\bibinfo{number}{3}), pp. \bibinfo{pages}{154--170},
  \doi{10.2307/2310058}.

\bibitemdeclare{article}{Lam:ded1}
\bibitem{Lam:ded1}
\bibinfo{author}{Joachim \surnamestart Lambek\surnameend}
  (\bibinfo{year}{1968}): \emph{\bibinfo{title}{Deductive Systems and
  Categories {I}: Syntactic Calculus and Residuated Categories}}.
\newblock {\sl \bibinfo{journal}{Mathematical Systems Theory}}
  \bibinfo{volume}{2}(\bibinfo{number}{4}), pp. \bibinfo{pages}{287--318},
  \doi{10.1007/bf01703261}.

\bibitemdeclare{incollection}{Lam:ded2}
\bibitem{Lam:ded2}
\bibinfo{author}{Joachim \surnamestart Lambek\surnameend}
  (\bibinfo{year}{1969}): \emph{\bibinfo{title}{Deductive Systems and
  Categories {II}: Standard Constructions and Closed Categories}}.
\newblock In: {\sl \bibinfo{booktitle}{Category theory, Homology Theory and
  Their Applications {I}}}, \bibinfo{publisher}{Springer}, pp.
  \bibinfo{pages}{76--122}, \doi{10.1007/bfb0079385}.

\bibitemdeclare{phdthesis}{Pol:ordlla}
\bibitem{Pol:ordlla}
\bibinfo{author}{Jeff \surnamestart Polakov\surnameend} (\bibinfo{year}{2001}):
  \emph{\bibinfo{title}{Ordered Linear Logic and Applications}}.
\newblock Ph.D. thesis, \bibinfo{school}{Carnegie Mellon University, {USA}}.
\newblock \urlprefix\url{https://www.cs.cmu.edu/~jpolakow/diss.ps}.

\bibitemdeclare{inproceedings}{PP:natdin}
\bibitem{PP:natdin}
\bibinfo{author}{Jeff \surnamestart Polakow\surnameend} \&
  \bibinfo{author}{Frank \surnamestart Pfenning\surnameend}
  (\bibinfo{year}{1999}): \emph{\bibinfo{title}{Natural Deduction for
  Intuitionistic Non-communicative Linear Logic}}.
\newblock In \bibinfo{editor}{Jean{-}Yves \surnamestart Girard\surnameend},
  editor: {\sl \bibinfo{booktitle}{Proceedings of the 4th International
  Conference on Typed Lambda Calculi and Applications, {TLCA} 1999}}, {\sl
  \bibinfo{series}{Lecture Notes in Computer Science}} \bibinfo{volume}{1581},
  \bibinfo{publisher}{Springer}, pp. \bibinfo{pages}{295--309},
  \doi{10.1007/3-540-48959-2\_21}.

\bibitemdeclare{article}{Roo:pronlc}
\bibitem{Roo:pronlc}
\bibinfo{author}{Dirk \surnamestart Roorda\surnameend} (\bibinfo{year}{1992}):
  \emph{\bibinfo{title}{Proof Nets for Lambek Calculus}}.
\newblock {\sl \bibinfo{journal}{Journal of Logic and Computation}}
  \bibinfo{volume}{2}(\bibinfo{number}{2}), pp. \bibinfo{pages}{211--231},
  \doi{10.1093/logcom/2.2.211}.

\bibitemdeclare{inproceedings}{UVZ:dedscs}
\bibitem{UVZ:dedscs}
\bibinfo{author}{Tarmo \surnamestart Uustalu\surnameend},
  \bibinfo{author}{Niccol{\'o} \surnamestart Veltri\surnameend} \&
  \bibinfo{author}{Noam \surnamestart Zeilberger\surnameend}
  (\bibinfo{year}{2021}): \emph{\bibinfo{title}{Deductive Systems and Coherence
  for Skew Prounital Closed Categories}}.
\newblock In \bibinfo{editor}{Claudio \surnamestart {Sacerdoti
  Coen}\surnameend} \& \bibinfo{editor}{Alwen \surnamestart Tiu\surnameend},
  editors: {\sl \bibinfo{booktitle}{Proceedings of the 15th Workshop on Logical
  Frameworks and Meta-Languages: Theory and Practice, {LFMTP} 2020}}, {\sl
  \bibinfo{series}{Electronic Proceedings in Theoretical Computer Science}}
  \bibinfo{volume}{332}, \bibinfo{publisher}{Open Publishing Assoc.}, pp.
  \bibinfo{pages}{35--53}, \doi{10.4204/eptcs.332.3}.

\bibitemdeclare{inproceedings}{VR:expebc}
\bibitem{VR:expebc}
\bibinfo{author}{Nachiappan \surnamestart Valliappan\surnameend} \&
  \bibinfo{author}{Alejandro \surnamestart Russo\surnameend}
  (\bibinfo{year}{2019}): \emph{\bibinfo{title}{Exponential Elimination for
  Bicartesian Closed Categorical Combinators}}.
\newblock In \bibinfo{editor}{Ekaterina \surnamestart
  Komendantskaya\surnameend}, editor: {\sl \bibinfo{booktitle}{Proceedings of
  the 21st International Symp. on Principles and Practice of Programming
  Languages, {PPDP} 2019}}, \bibinfo{publisher}{{ACM}}, pp.
  \bibinfo{pages}{20:1--20:13}, \doi{10.1145/3354166.3354185}.

\end{thebibliography}

\end{document}